# LASER EXCITATION OF WAKEFIELDS IN A PARABOLIC PLASMA CHANNEL

R. Annou[1, 3], V. K. Tripathi [1], R. K. Jarwal[1], and A. K. Sharma [2]

[1] Physics Department,
[2] Center for Energy Studies,
Indian Institute of Technology, Delhi,
New Delhi 110016, INDIA.
[3] Permanent address: Faculty of Physics, USTHB, ALGERIA

**Abstract.** A Gaussian laser pulse with a square profile in time propagating in a parabolic plasma channel, exerts a ponderomotive force on electrons that drives a large amplitude electrostatic wake potential $\phi$. An analytic solution for $\phi$ is possible when a characteristic relation is fulfilled. For pulse duration $\tau \sim 9\ \tau_p / 4$, where $\tau_p$ is the plasma period, the plasma wake potential on the axis is twice the ponderomotive potential $\phi_0$.

**Keywords:** Plasma channels, acceleration, wake fields.

Laser produced plasmas may be utilized for charged particles acceleration purpose. The plasma beat-wave, where two laser beams of frequency difference equal to the plasma frequency are used to drive resonantly a plasma wave, and the wakefield accelerator where, a single pulse is employed to excite the large amplitude plasma wave, are two major concepts for acceleration. Electron acceleration to energies up to ~ 3 GeV has been achieved by exciting a large amplitude plasma wave by the beat frequency ponderomotive force due to two lasers, or to a single laser pulse [1,2]. The maximum energy acquired depends linearly on the interaction length which is times the Rayleigh length in optical guiding case. The guiding of the pump lasers to long distance is required [3] and a preformed plasma channel is considered to be useful. Durfee III *et al* [4] have indeed succeeded to guide a pulse over ~24 Rayleigh lengths in a channel already formed by an other pulse (pre-pulse) .The excitation of a plasma wave by beating two lasers in a plasma channel with parabolic density profile has been then analyzed by Sharma *et al* in Ref.[5].

In this letter, we study analytically the excitation of wakefields by a laser pulse in a parabolic density profile plasma channel.

Consider a laser pulse of frequency $\omega_0$ propagating in a plasma channel with a parabolic density profile $\omega_p^2 = \omega_{p0}^2\ (1 + x^2 / L^2)$. The evolution of the wake potential is governed by the following set of equations,

$$\frac{\partial \vec{v}}{\partial t} = \frac{e}{m}\nabla(f + f_p), \tag{1}$$

$$\frac{\partial n}{\partial t} + \nabla \cdot (n_0 \vec{v}) = 0, \tag{2}$$

$$\Delta\ f = 4 p\, e\, n\ , \tag{3}$$

where, $f_p = f_0\, g(t)\, \exp(-\, x^2 / x_0^2)$ is the ponderomotive potential, $f_0 = -\frac{1}{2e} m_e v_{osc}^2$, $v_{osc}$ being the oscillatory velocity.

By Laplace transforming and recombining Eqs.(1-3) , one obtains,

$$\Delta\ \tilde{f} + \frac{\nabla w_p^2\ .\ \nabla \tilde{f}}{p^2 + w_p^2} = y\,(p, x)\ , \tag{4}$$

where, $y\,(p, x) = -\frac{\nabla w_p^2 \cdot \nabla \tilde{f}_p}{p^2 + w_p^2}$ , $p = i w$, and ~ stands for Laplace transform .

By considering x and z dimensions, where the z dependence is written as $\tilde{f} = \tilde{f}\,\sin(k_z z)$ Eq.(4) reduces to ,

$$(1 - l^2)\frac{\partial^2 \tilde{f}}{\partial l^2} - 2\,l\,\frac{\partial \tilde{f}}{\partial l} + (1 - l^2)\,g\,\tilde{f} = \bar{y}\,(p, l)\ , \tag{5}$$

where, $x = i\lambda\alpha$, $a^2 = L^2(1 + p^2 / w_{p0}^2)$, $g = k_z^2\, a^2$ and $\bar{y}(p, l) = -\,a^2\ y(p, i\,l\,a)$ .

Without the RHS, Eq.(5) reduces to a spheroidal wave equation. And for a solution of the homogeneous equation near the origin to exist, $g$ must fulfill the following condition (c.f. Ref.[6]),

$$0=\gamma+\frac{1\times2g}{g-2\times3+\frac{3\times4g}{g-4\times5+...}}=\gamma+\frac{1\times2g}{g-2\times3}+.. \quad (6)$$

It is clear from Eq.(6) that $w=w_{p0}$ is a solution.

This condition constitutes the dispersion relation for the eigen modes of the parabolic plasma channel (plasma fiber). Whereas the solutions valid in the segment (-1, 1) are,

$$\tilde{f}\equiv(\ Q^0_{s0}(l,g)=\sum_{-\infty}^{\infty}(-1)^r a^0_{0,r}(g)\ Q^0_{2r}(l);$$

$$P^0_{s0}(l,g)=\sum_{-\infty}^{\infty}(-1)^r a^0_{0,r}(g)\ P^0_{2r}(l)\ ),$$

where, $Q^0_{2r}(l)$ and $P^0_{2r}(l)$ are associated Legendre polynomials.

Moreover, putting $\tilde{f}=\frac{y}{\sqrt{1-l^2}}$, leads to

$$y''+\left[g+\frac{1}{(1-l^2)^2}\right]y=\bar{y}(p,l)\sqrt{1-l^2} \quad (7)$$

and solution of Eq.(7) is given by

$$y(l)=\tilde{f}\sqrt{1-l^2}=\sqrt{1-l^2}\ P^0_{s0}(l,g)\int\frac{\int(1-l^2)\bar{y}(p,l)\,dl}{(1-l^2)P^0_{s0}(l,g)}dl \quad (8)$$

where, $\sqrt{1-l^2}\ P^0_{s0}(l,g)$ is the solution for the homogeneous equation corresponding to Eq.(75). On the axis the relation (8) reduces to ,

$$\frac{\tilde{f}(p,0)}{f_0}=\tilde{g}(t)\frac{w^2_{p0}}{p^2+w^2_{p0}}\ \sin(k_z z) \quad (9)$$

For a square pulse in time, i.e., g(t) = u(t) - u(t-τ) , where u(t) is the step function and τ is the pulse duration, one obtains,

$$\frac{f}{f_0}=(1-\cos\frac{2p\,t}{t_p}-[1-\cos\frac{2p(t-t)}{t_p}]u(t-t))\sin(k_z z). \quad (10)$$

where, $t_p=2p/w_{p0}$. Furthermore, when the pulse is gone, the wake potential is expressed as follows,

$$\frac{f(t,0,z)}{f_0}=2\sin(2p\frac{t-t/2}{t_p})\ \sin\frac{2pt}{t_p}\ \sin(k_z z), \quad (11)$$

To have an effective excitation we may choose $\tau/\tau_p = 9/4$, and the maximum attainable is $\phi_{max}/\phi_0=2$.

To conclude, we recall that the excitation of wakefields by a Gaussian laser pulse with a square temporal profile, is analytically investigated and exact solution is found in a plasma channel[7]. The density profile in the channel is taken parabolic; which is a good approximation to the real profile in the paraxial region, however it ceases to be valid elsewhere as the density grows infinitely (a more realistic profile has been studied in Ref.[8]).The wake potential is found to be periodic in terms of the pulse duration τ, and the maximum magnitude attainable on the axis is twice that of the ponderomotive potential.

## ACKNOWLEDGMENTS

R.A, acknowledges the support of USTHB (ALGERIA)/ DST-INDIA.